\documentclass[12pt]{article}
\usepackage{amsfonts,amssymb}

\newcommand{\R}[1]{\mathbb{R}^{#1}}

\newcommand{\beq}{\begin{eqnarray}}
\newcommand{\eeq}{\end{eqnarray}}

\def \bi{\bibitem}
\def\){\right)}
\def\({\left( }

\def\tr{{\rm\ Tr}}

\newcommand{\vev}[1]{\langle#1\rangle}
\def\be{\begin{equation}}
\def\ee{\end{equation}}
\def\bea{\begin{eqnarray}}
\def\eea{\end{eqnarray}}
\newcommand{\eref}[1]{(\ref{#1})}

\newcommand{\rem}[1]{}
\def\tr{{\rm tr}}

\def\NN{{\cal N}}

\def\none{$\NN=1$}

\def\ZZ{{\mathbb Z}}
\def\S{{\bf S}}

\def\RR{{\bf R}}

\def\N{{\bf N}}
\def\Tr{{\rm Tr}}

\def \bi{\bibitem}
\def\np {  {\em Nucl. Phys.} }

\def \pr  { {\em Phys. Rev.} }

\def\ltap{\ \raise.3ex\hbox{$<$\kern-.75em\lower1ex\hbox{$\sim$}}\ }
\def\gtap{\ \raise.3ex\hbox{$>$\kern-.75em\lower1ex\hbox{$\sim$}}\ }

\title{D-Branes on the Conifold
and ${\cal N}=1$ Gauge/Gravity Dualities\footnote{
Based on I.~R.~K.'s lectures at the Les Houches Summer School
Session 76, ``Gravity, Gauge Theories, and Strings'', August 2001.}}
\author{Christopher P. Herzog, Igor R. Klebanov and
Peter Ouyang\\
Department of Physics, Princeton University\\
Princeton, NJ 08544, USA\\
  }
\begin{document}
\setlength{\baselineskip}{16pt}
\begin{titlepage}
\maketitle
\begin{picture}(0,0)(0,0)
\put(325,245){PUPT-2039}
\put(325,260){hep-th/0205100}
\end{picture}
\vspace{-36pt}
\begin{abstract}

We review extensions of the AdS/CFT correspondence
to gauge/ gravity dualities with ${\cal N}=1$ supersymmetry.
In particular, we describe
the gauge/gravity dualities that emerge from placing
D3-branes at the apex of the conifold.
We consider first the conformal case,
with discussions
of chiral primary operators and wrapped D-branes.
Next, we break the conformal symmetry by
adding a stack of partially wrapped D5-branes
to the system, changing the gauge group
and introducing
a logarithmic renormalization group flow.
In the gravity dual, the effect
of these wrapped D5-branes is to turn on
the flux of 3-form field strengths.  The associated 
RR 2-form potential breaks the $U(1)$ R-symmetry 
to $\ZZ_{2M}$ and we study this phenomenon in detail.  
This extra flux also leads to deformation
of the cone near the apex,
which describes the chiral symmetry breaking and confinement
in the dual gauge theory.
\end{abstract}
\thispagestyle{empty}
\setcounter{page}{0}
\end{titlepage}

\section{Introduction}

Comparison of a stack of D3-branes with the geometry it produces
leads to formulation of duality between ${\cal N}=4$ supersymmetric
Yang-Mills theory and type II strings on $AdS_5\times \S^5$
\cite{jthroat,US,EW}. It is of obvious interest to consider more
general dualities between gauge theories and string theories
where some of the supersymmetry and/or conformal invariance are broken.
These notes are primarily devoted to extensions of the AdS/CFT
correspondence  to theories with ${\cal N}=1$
supersymmetry.

We first show how to break some of the supersymmetry
without destroying conformal invariance.  This may be
accomplished through placing a stack of D3-branes at the apex of a
Ricci flat 6-dimensional cone \cite{ks,Kehag,KW,MP}.  Then we show how
to break the conformal invariance in this set-up and to introduce
logarithmic RG flow into the field theory.  A convenient way to make
the coupling constants run logarithmically is to introduce fractional
D3-branes at the apex of the cone \cite{GK,KN,KT}; these fractional
branes may be thought of as D5-branes wrapped over 2-cycles in the
base of the cone.  In the gravity dual the effect of these wrapped
D5-branes is to turn on the flux of 3-form field strengths. This extra
flux may lead to deformation of the cone near the apex, which
describes the chiral symmetry breaking and confinement in the dual
gauge theory \cite{KS}.  We will start the notes with a very brief
review of some of the basic facts about the AdS/CFT
correspondence. For more background the reader may consult, for
example, the review papers \cite{magoo,me}.

To make the discussion more concrete, we consider primarily one
particular example of a cone, the conifold.  There are two reasons for
this focus.  The conifold has enough structure that many new aspects
of AdS/CFT correspondence emerge that are not immediately visible for
the simplest case, where the conifold is replaced with $\R{6}$.  At
the same time, the conifold is simple enough that we can follow the
program outlined in the paragraph above in great detail.
This program eventually leads to the warped
deformed conifold \cite{KS}, a solution of type IIB supergravity that
is dual to a certain ${\cal N}=1$ supersymmetric $SU(N+M)\times SU(N)$
gauge theory in the limit of strong `t Hooft coupling. This solution
encodes various interesting gauge theory phenomena in a dual
geometrical language, such as
the chiral anomaly, the logarithmic running of couplings,
the duality cascade in the UV, and chiral
symmetry breaking and confinement in the IR.

First, however, we review the original AdS/CFT correspondence. The
duality between ${\cal N}=4$ supersymmetric $SU(N)$ gauge theory and
the $AdS_5\times \S^5$ background of type IIB string theory
\cite{jthroat,US,EW} is usually motivated by considering a stack of a
large number $N$ of D3-branes. The SYM theory is the low-energy limit
of the gauge theory on the stack of D3-branes.
On the other hand, the curved
background produced by the stack is
\be
\label{geom}
ds^2 = h^{-1/2}
\left (- dt^2 +dx_1^2+ dx_2^2+ dx_3^2\right )
+ h^{1/2}
\left ( dr^2 + r^2 d\Omega_5^2 \right )\ ,
\ee
where $d\Omega_5^2$ is the metric of a unit 5-sphere and
\be
h(r) = 1+{L^4\over r^4}\ .
\ee
This 10-dimensional metric may be thought of as a ``warped product''
of the $\R{3,1}$ along the branes and the transverse space $\R{6}$.
Note that the dilaton, $\Phi=0$, is constant, and
the selfdual 5-form field strength is given by
\be \label{firstfive}
F_5 = {\cal F}_5 + \star {\cal F}_5 \ , \qquad  {\cal F}_5
=16 \pi (\alpha')^2 N {\rm vol}(\S^5)
\ .
\ee
The normalization above is dictated by the
quantization of  D$p$-brane tension which implies
\be \label{quantten}
\int_{\S^{8-p}} \star F_{p+2} = {2\kappa^2 \tau_p N \over g_s}
\ ,
\ee
where
\be
\tau_p = {\sqrt\pi\over \kappa} (4\pi^2\alpha')^{(3-p)/2}
\ee
and $\kappa = 8 \pi^{7/2} g_s \alpha'^2$ is the 10-dimensional
gravitational constant.
In particular, for $p=3$ we have
\be
\int_{\S^5} F_5 = (4\pi^2 \alpha')^2 N
\ ,\ee
which is consistent with (\ref{firstfive})
since the volume of a unit 5-sphere is 
\[
{\rm Vol}(\S^5)=\pi^3 \ .
\]

Note that the 5-form field strength may also be written as
\be
g_s F_5 = d^4 x\wedge d h^{-1} - r^5 {dh\over dr} {\rm vol}(\S^5) \ .
\ee
Then it is not hard to see that
the Einstein equation
\[
R_{MN} = \frac{g_s^2}{96} F_{MPQRS} {F_N}^{PQRS} 
\]
is satisfied.
Since $-r^5 {dh\over dr} = 4 L^4$, we find
by comparing with (\ref{firstfive}) that
\be
L^4 = 4\pi g_s N \alpha'^2\ .
\ee

A related way to determine the scale factor $L$ is to equate the ADM
tension of the supergravity solution with $N$ times the tension of a
single D3-brane \cite{gkp}:
\be
{2\over \kappa^2} L^4 {\rm Vol}(\S^5) = {\sqrt \pi\over \kappa} N
\ .
\ee
This way we find
\be
L^4 = {\kappa N\over 2\pi^{5/2}} = 4\pi g_s N \alpha'^2\
\ee
in agreement with the preceding paragraph.

The radial coordinate $r$ is related to the scale in the dual gauge
theory.  The low-energy limit corresponds to $r\rightarrow 0$.  In
this limit the metric becomes
\be \label{adsmetric}
ds^2 = {L^2 \over z^2} \left( -dt^2 + d\vec{x}^2 + dz^2 \right) +
    L^2 d\Omega_5^2 \ ,
\ee
where $z={L^2\over r}$. This describes the direct product of
5-dimensional Anti-de Sitter space, $AdS_5$, and the 5-dimensional
sphere, ${\bf S}^5$, with equal radii of curvature $L$.

An interesting generalization of the basic AdS/CFT correspondence
\cite{jthroat,US,EW} is found by studying branes at conical
singularities \cite{ks,Kehag,KW,MP}.  Consider a stack of D3-branes
placed at the apex of a Ricci-flat 6-d cone $Y_6$ whose base is a 5-d
Einstein manifold $X_5$. Comparing the metric with the D-brane
description leads one to conjecture that type IIB string theory on
$AdS_5\times X_5$ is dual to the low-energy limit of the world volume
theory on the D3-branes at the singularity. The equality of tensions
now requires \cite{Gubser}
\be \label{radius}
L^4 = {\sqrt \pi \kappa N\over 2 {\rm Vol}(X_5) } = 4\pi g_s N \alpha'^2
{\pi^3\over {\rm Vol}(X_5) } \ ,
\ee
an important normalization formula which we will use in the following
section.

The simplest examples of $X_5$ are the orbifolds ${\bf S}^5/\Gamma$
where $\Gamma$ is a discrete subgroup of $SO(6)$ \cite{ks}. In these
cases $X_5$ has the local geometry of a 5-sphere.  The dual gauge
theory is the IR limit of the world volume theory on a stack of $N$
D3-branes placed at the orbifold singularity of $\R{6}/\Gamma$. Such
theories typically involve product gauge groups $SU(N)^k$ coupled to
matter in bifundamental representations \cite{dm}.

Constructions of the dual gauge theories for Einstein manifolds $X_5$
which are not locally equivalent to ${\bf S}^5$ are also possible.
The simplest example is the Romans compactification on $X_5= T^{1,1}=
(SU(2)\times SU(2))/U(1)$ \cite{Romans,KW}.  The dual gauge theory is
the conformal limit of the world volume theory on a stack of $N$
D3-branes placed at the singularity of a Calabi-Yau manifold known as
the conifold \cite{KW}, which is a cone over $T^{1,1}$. Let us explain
this connection in more detail.


\section{D3-branes on the Conifold}

The conifold may be described by the
following equation in four complex variables,
\be \label{coni}
\sum_{a=1}^4 z_a^2 = 0
\ .
\ee
Since this equation is invariant under an overall real rescaling of
the coordinates, this space is a cone. Remarkably, the base of this
cone is precisely the space $T^{1,1}$ \cite{cd,KW}. In fact, the
metric on the conifold may be cast in the form \cite{cd}
\be
ds_6^2 = dr^2 + r^2 ds_{T^{1,1}}^2\ ,
\label{conimetric}
\ee
where
\begin{equation} \label{co}
ds_{T^{1,1}}^2=
{1\over 9} \bigg(d\psi +
\sum_{i=1}^2 \cos \theta_i d\phi_i\bigg)^2+
{1\over 6} \sum_{i=1}^2 \left(
d\theta_i^2 + {\rm sin}^2\theta_i d\phi_i^2
 \right)
\ 
\end{equation}
is the metric on $T^{1,1}$. Here $\psi$ is an angular coordinate
which  ranges from $0$ to $4\pi$,  while $(\theta_1,\phi_1)$
and $(\theta_2,\phi_2)$ parametrize two ${\bf S}^2$s in a standard way.
Therefore, this form of the metric shows that
$T^{1,1}$ is an ${\bf S}^1$ bundle over ${\bf S}^2 \times {\bf S}^2$.

Now placing $N$ D3-branes at the apex of the cone we find the metric
\begin{eqnarray}
ds^2 &=& \left (1+{L^4\over r^4}\right )^{-1/2}
\left (- dt^2 +dx_1^2+ dx_2^2+ dx_3^2\right ) \nonumber
\\
&& + \left (1+{L^4\over r^4}\right )^{1/2}
\left ( dr^2 + r^2 ds_{T^{1,1}}^2 \right )
\label{newgeom}
\end{eqnarray}
whose near-horizon limit is $AdS_5\times T^{1,1}$.
Using the metric (\ref{co}) it is not hard to find that the
volume of $T^{1,1}$ is ${16 \pi^3 \over 27}$
\cite{GK}. From (\ref{radius}) it then follows that
\be \label{quantiz}
L^4 = 4\pi g_s N (\alpha')^2 {27\over 16} =
{27 \kappa N\over 32\pi^{5/2}}
\ .
\ee
The same logic that leads us to the maximally supersymmetric
version of the AdS/CFT correspondence now shows that
the type IIB string theory on this space should
be dual to the infrared limit of the field theory on $N$ D3-branes
placed at the singularity of the conifold. Since Calabi-Yau spaces
preserve 1/4 of the original supersymmetries we find that this
should be an ${\cal N}=1$
superconformal field theory.
This field theory was constructed
in \cite{KW}: it is $SU(N)\times SU(N)$ gauge theory
coupled to two chiral superfields, $A_i$, in the
$({\bf N}, \overline{\bf N})$
representation
and two chiral superfields, $B_j$, in the $(\overline{\bf N}, {\bf N})$
representation. The $A$'s transform as a doublet under one
of the global $SU(2)$s while the $B$'s transform
as a doublet under the other $SU(2)$.

A simple way to motivate the appearance of the fields $A_i,\ B_j$ is to
rewrite the defining equation of the conifold, (\ref{coni}), as
\begin{equation} \label{conifold}
\det_{i,j} z_{ij} = 0\ ,
\qquad z_{ij} ={1\over \sqrt{2}}\sum_n \sigma^n_{ij} z_n
\end{equation}
where $\sigma^n$ are the Pauli matrices for $n=1,2,3$
and $\sigma^4$ is $i$ times the unit matrix.
This quadratic constraint may be ``solved'' by the substitution
\begin{equation}
z_{ij} = A_i B_j
\ ,
\end{equation}
where $A_i,\ B_j$ are unconstrained variables.
If we place a single D3-brane at the singularity
of the conifold, then we find
a $U(1)\times U(1)$ gauge theory coupled to fields $A_1, A_2$ with
charges $(1, -1)$ and $B_1, B_2$ with charges $(-1,1)$.

In constructing the generalization to the non-abelian theory on
$N$ D3-branes,
cancellation of the anomaly in the $U(1)$ R-symmetry requires that
the $A$'s and the $B$'s each have R-charge $1/2$. For consistency of
the duality it is necessary that we add
an exactly marginal superpotential which preserves the
$SU(2)\times SU(2)\times U(1)_R$ symmetry of the theory (this
superpotential produces a critical line related to the radius of
$AdS_5\times T^{1,1}$). Since a marginal superpotential has R-charge
equal to 2 it must be quartic, and the symmetries fix it uniquely
up to overall normalization:
\be \label{superpotential}
W=\epsilon^{ij}
\epsilon^{kl}\tr A_iB_kA_jB_l\ .
\ee
Therefore, it was proposed in \cite{KW} that the $SU(N)\times
SU(N)$ SCFT with this superpotential is dual to type IIB strings
on $AdS_5\times T^{1,1}$.

This proposal can be checked in an interesting
way by comparing to a certain
$AdS_5\times {\bf S}^5/\ZZ_2$ background.  If ${\bf S}^5$ is described
by an equation
\be\label{beqnop}
\sum_{i=1}^6x_i^2=1,
\ee
with real variables $x_1,\dots, x_6$,
then the $\ZZ_2$ acts as $-1$ on four of the $x_i$ and
as $+1$ on the other two.  The importance of this choice is that this
particular
 $\ZZ_2$ orbifold of $AdS_5\times {\bf S}^5$
  has ${\cal N}=2$ superconformal symmetry.
Using orbifold results for D-branes \cite{dm}, this model has been
identified \cite{ks} as an AdS dual of a $U(N)\times U(N)$ theory
with hypermultiplets transforming in
$(\N,\overline\N)\oplus (\overline \N,\N)$. From an ${\cal N}=1$ point of
view, the hypermultiplets
correspond to chiral multiplets $A_k,B_l$, $k,l=1,2$ in the
$(\N,\overline \N)$ and $(\overline \N,\N)$  representations respectively.
The model also contains, from an ${\cal N}=1$ point of view, chiral multiplets
$\Phi$ and $\tilde \Phi$ in the adjoint representations of the two
$U(N)$'s.
The superpotential is
$$ g \Tr \Phi (A_1 B_1 - A_2 B_2) + g \Tr \tilde \Phi (B_1 A_1 - B_2 A_2)
\ .
$$
Now, let us add to the superpotential of this $\ZZ_2$ orbifold
a relevant term,
\be \label{relper}{m\over 2} (\Tr \Phi^2 - \Tr \tilde \Phi^2 )
\ .
\ee
It is straightforward to see what this does to the field theory.
We simply integrate out $\Phi$ and $\tilde \Phi$,
to find the superpotential
$$ -{g^2\over m} \left [\Tr (A_1 B_1 A_2 B_2) - \Tr (B_1 A_1 B_2 A_2)
\right ]\ .
$$
This expression is the same as (\ref{superpotential}), so the
$\ZZ_2$ orbifold with relevant perturbation (\ref{relper})
apparently flows to the $T^{1,1}$ model associated with the conifold.

Let us try to understand why this works from the point of view of the
geometry of ${\bf S}^5/\ZZ_2$.  The perturbation in (\ref{relper})
is odd under exchange of the 
two $U(N)$'s.  The exchange of the two $U(N)$'s
is the quantum symmetry of the $AdS_5\times {\bf S}^5/\ZZ_2$ orbifold
-- the symmetry that acts as $-1$ on string states in the twisted sector
and $+1$ in the untwisted  sector.  Therefore we
associate this perturbation with a twisted sector mode of string
theory on $AdS_5\times {\bf S}^5/\ZZ_2$.  The twisted sector mode
which is a relevant perturbation of the field theory is the blowup
of the orbifold singularity of ${\bf S}^5/\ZZ_2$ into the smooth space
$T^{1,1}$. A somewhat different derivation of the field theory
on D3-branes at the conifold singularity, which is based on blowing
up a $\ZZ_2\times \ZZ_2$ orbifold, was given in \cite{MP}.

It is interesting to examine how various quantities change under
the RG flow from the ${\bf S}^5/\ZZ_2$ theory to the
$T^{1,1}$ theory. The behavior of the conformal anomaly (which is equal to
the $U(1)^3_R$ anomaly) was studied in \cite{Gubser}. Using the fact
that the chiral superfields carry
R-charge equal to $1/2$, on the field theory
side it was found that
\be \label{anom}
{c_{IR}\over c_{UV}} = {27\over 32}
\ .
\ee
On the other hand, all 3-point functions calculated from supergravity on
$AdS_5\times X_5$ carry normalization factor inversely proportional
to ${\rm Vol}(X_5)$. Thus, on the supergravity side
\be
{c_{IR}\over c_{UV}} = {{\rm Vol}\ ({\bf S}^5/\ZZ_2)
\over {\rm Vol}\ (T^{1,1}) }
= {27\over 32}\ .
\ee
Thus, the supergravity calculation
is in exact agreement with the field theory result
(\ref{anom}) \cite{Gubser}.
This is a striking and highly sensitive test of the ${\cal N}=1$
dual pair constructed in \cite{KW,MP}.

\subsection{ Dimensions of Chiral Operators}

There are a number of further convincing checks of the duality between
this field theory and type IIB strings on $AdS_5\times T^{1,1}$.
Here we discuss the supergravity modes
which correspond to chiral primary operators. (For a more extensive
analysis of the spectrum of the model, see \cite{Ceres}.)
For the $AdS_5\times {\bf S}^5$
case, these modes are mixtures of the conformal factors of the
$AdS_5$ and ${\bf S}^5$ and the 4-form field.
The same has been shown to be true for the
$T^{1,1}$ case \cite{Gubser,RD,Ceres}.
In fact, we may keep the discussion of such modes quite
general and consider $AdS_5\times X_5$ where $X_5$ is any Einstein manifold.

The diagonalization of such modes carried out in \cite{Kim}
for the ${\bf S}^5$ case is easily generalized to any $X_5$.
The mixing of the conformal factor and 4-form modes results in
the following mass-squared matrix,
\be m^2 = \pmatrix{ E+32 &  8E\cr  4/5 & E\cr}
\ee
where $E\geq 0$ is the eigenvalue of the Laplacian on $X_5$.
The eigenvalues of this matrix are
\be \label{masses}
 m^2 = 16 + E \pm 8 \sqrt{ 4+E}
\ .
\ee

We will be primarily interested in the modes which correspond
to picking the minus branch: they turn out to be the chiral primary
fields. For such modes there is a possibility of $m^2$ falling in
the range
\be \label{range}
-4 < m^2 < - 3
\ee 
where there is a two-fold ambiguity in defining
the corresponding operator dimension \cite{KWnew}. 

First, let us recall the ${\bf S}^5$ case where the spherical
harmonics correspond to
traceless symmetric tensors of $SO(6)$, $d^{(k)}_{i_1\ldots i_k}$.
Here $E= k(k+4)$, and it seems that
the bound (\ref{range}) is satisfied for $k=1$. However, this is precisely the
special case where the corresponding mode is missing:
for $k=1$ one of the two mixtures is the singleton \cite{Kim}.
Thus, all chiral primary
operators in the ${\cal N}=4$ $SU(N)$ theory correspond to
the conventional branch of dimension, $\Delta_+$.
It is now well-known that this family of operators with dimensions
$\Delta= k$, $k=2,3,\ldots$
is $d^{(k)}_{i_1\ldots i_k} \Tr (X^{i_1} \ldots X^{i_k})$.
The absence of $k=1$ is related to the gauge group being $SU(N)$ rather
than $U(N)$. Thus, in this case we do not encounter operator
dimensions lower than $2$.

The situation is different for $T^{1,1}$. Here there is a family of wave
functions labeled by non-negative integer $k$, transforming under
$SU(2)\times SU(2)$  as $(k/2,k/2)$, and with
$U(1)_R$ charge $k$ \cite{Gubser,RD,Ceres}.
The corresponding eigenvalues of the Laplacian
are
\be \label{honeypot}
E(k)=3\left(k(k+2)
-{k^2\over 4}\right)\
.
\ee
In \cite{KW} it was argued that the dual chiral operators are
\be \label{chirop}
\tr (A_{i_1} B_{j_1} \ldots A_{i_k} B_{j_k} )
\ .
\ee
Since the F-term constraints in the gauge theory require that the
$i$ and the $j$ indices are separately symmetrized, we find that
their $SU(2)\times SU(2)\times U(1)$ quantum numbers agree with those
given by the supergravity analysis. In the field theory
the $A$'s and the $B$'s have dimension $3/4$, hence the dimensions
of the chiral operators are $3k/2$.

In studying the dimensions from the supergravity point of view, one encounters
an interesting subtlety discussed in \cite{KWnew}. While for $k>1$ only
the dimension $\Delta_+$ is admissible, for $k=1$ one could pick either
branch. Indeed, from (\ref{honeypot}) we have
$E(1)=33/4$ which falls within the range (\ref{range}). Here we
find that $\Delta_-=3/2$, while $\Delta_+=5/2$. Since the supersymmetry
requires the corresponding dimension to be $3/2$, in this case
we have to pick the unconventional $\Delta_-$
branch \cite{KWnew}. Choosing this
branch for $k=1$ and $\Delta_+$ for $k>1$ we indeed find
following \cite{Gubser,RD,Ceres}
that the supergravity analysis based on (\ref{masses}), (\ref{honeypot})
reproduces the dimensions $3k/2$ of the chiral operators (\ref{chirop}).
Thus, the conifold theory provides
a simple example of AdS/CFT duality where the $\Delta_-$ branch
has to be chosen for certain operators.

Let us also note that substituting $E(1)=33/4$ into (\ref{masses}) we find
$m^2=-15/4$ which corresponds to a conformally coupled scalar in $AdS_5$
\cite{Kim}. In fact, the short chiral
supermultiplet containing this scalar includes
another conformally coupled scalar and a massless fermion \cite{Ceres}.
One of these
scalar fields corresponds to the lower component of the superfield
$\Tr (A_i B_j)$, which has dimension $3/2$, while the other
corresponds to the upper component which has dimension $5/2$. Thus,
the supersymmetry requires that we pick dimension $\Delta_+$ for one
of the conformally coupled scalars, and $\Delta_-$ for the other.

\subsection{Wrapped D3-branes as ``dibaryons''}

It is of further interest to consider various branes wrapped
over the cycles of $T^{1,1}$ and attempt to
identify these states in the field theory \cite{GK}.
For example, wrapped
D3-branes turn out to correspond to baryon-like operators $A^N$
and $B^N$ where the indices of both $SU(N)$ groups are fully
antisymmetrized.  For large $N$ the dimensions of such operators
calculated from the supergravity are found to be $3N/4$ \cite{GK}.
This is in
complete agreement with the fact that the dimension of the chiral
superfields at the fixed point is $3/4$ and may be regarded as a direct
supergravity calculation of an anomalous dimension in the dual gauge
theory.

To show how this works in detail, we need to calculate the mass
of a D3-brane wrapped over a minimal volume 3-cycle.
An example of such a 3-cycle is
the subspace at a constant value of $(\theta_2, \phi_2)$, and its volume
is found to be $V_3= 8\pi^2 L^3/9$ \cite{GK}.
The mass of the D3-brane
wrapped over the 3-cycle is, therefore,
 \be
m= V_3 {\sqrt\pi\over \kappa} = {8 \pi^{5/2} L^3\over 9\kappa}
\ .
\ee
For large $mL$, the corresponding operator dimension
$\Delta$ approaches
\be
m L=
{8 \pi^{5/2} L^4\over 9\kappa} = {3\over 4} N
\ ,
\ee
where in the last step we used (\ref{quantiz}).

Let us construct the corresponding operators in the dual gauge theory.
Since the fields $A^{\alpha}_{k\beta}$, $k=1,2$,
carry an index $\alpha$ in the $\N$ of $SU(N)_1$ and an index $\beta$
in the $\overline{\N}$ of $SU(N)_2$,
we can construct color-singlet ``dibaryon'' operators
by antisymmetrizing completely with respect to both groups:
\be\label{BaryonOne}
{\cal B}_{1 l}= \epsilon_{\alpha_1 \ldots \alpha_N}
\epsilon^{\beta_1\ldots \beta_N} D_{l}^{k_1\ldots k_N}
\prod_{i=1}^N A^{\alpha_i}_{k_i  \beta_i}
\ ,
\ee
where $D_l^{k_1\ldots k_N}$ is the completely
symmetric $SU(2)$ Clebsch-Gordon
coefficient corresponding to forming the ${\bf N+1}$ of $SU(2)$ out of $N$ 2's.
Thus the $SU(2)\times SU(2)$ quantum numbers of ${\mathcal B}_{1 l}$ are
$({\bf N+1}, {\bf 1})$. Similarly, we can construct ``dibaryon'' operators
which transform as $({\bf 1}, {\bf N+1} )$,
\be\label{BaryonTwo}
{\cal B}_{2 l}= \epsilon^{\alpha_1 \ldots \alpha_N}
\epsilon_{\beta_1\ldots \beta_N} D_{l}^{k_1\ldots k_N}
\prod_{i=1}^N B^{\beta_i}_{k_i  \alpha_i}
\ .
\ee
Under the duality these operators map to D3-branes classically localized at
a constant $(\theta_1,\phi_1)$. Thus, the existence of two types
of ``dibaryon'' operators is related on the supergravity
side to the fact that the base of the $U(1)$
bundle is ${\bf S}^2\times {\bf S}^2$. At the quantum level, the collective
coordinate for the wrapped D3-brane has to be quantized, and
this explains its $SU(2)\times SU(2)$ quantum numbers \cite{GK}.
The most basic check on the operator
identification is that, since the exact dimension of the $A$'s and the $B$'s
is $3/4$, the dimension of the ``dibaryon'' operators agrees exactly
with the supergravity calculation.

\subsection{Other ways of wrapping D-branes over cycles of $T^{1,1}$}

There are many other admissible ways of wrapping
branes over cycles of $T^{1,1}$ (for a complete list,
see \cite{Mukhi}). For example, a D3-brane may be
wrapped over a 2-cycle, which produces a string in $AdS_5$.
The tension of such a ``fat'' string
scales as
$L^2/\kappa \sim N (g_s N)^{-1/2}/\alpha'$. The non-trivial
dependence of the tension on the 't~Hooft coupling $g_s N$ indicates
that such a string is not a BPS saturated object. This should be
contrasted with the tension of a BPS string obtained in \cite{Ed}
by wrapping a
D5-brane over ${\bf RP}^4$: $T\sim N/\alpha'$.

In discussing wrapped 5-branes, we will limit explicit statements
to D5-branes: since a $(p,q)$ 5-brane is an $SL(2,\ZZ)$
transform of a D5-brane, our discussion may be
generalized to wrapped $(p,q)$ 5-branes using the $SL(2,\ZZ)$
symmetry of the Type IIB string theory.
If a D5-brane is wrapped over the entire $T^{1,1}$ then, according to
the arguments in \cite{Ed,GO}, it serves as a vertex connecting $N$
fundamental strings. Since each string ends on a charge in the
fundamental representation of one of the $SU(N)$'s, the resulting
field theory state is a baryon built out of external quarks.

If a
D5-brane is wrapped over an ${\bf S}^3$, with its remaining
two dimensions parallel to $\RR^{3,1}$, then we find a domain wall
in the dual field theory.
Consider positioning a ``fat'' string made of a wrapped D3-brane
orthogonally to the domain wall. As the string is brought through the
membrane, a fundamental string
stretched between them
is created. The origin of this effect
is creation of fundamental strings by crossing D5
and D3 branes, as shown in \cite{bdg,dfk}.

We should note, however, that the domain wall positioned at some
arbitrary $AdS_5$ radial coordinate $r$ is not stable: its energy
scales as $r^3$. Therefore, the only stable position is at $r=0$
which is the horizon. The domain wall is tensionless there, and it
is unlikely that this object really exists in the dual CFT.
We will see, however, that the domain wall made of a wrapped D5-brane
definitely exists in the $SU(N)\times SU(N+M)$ generalization of the
gauge theory. This theory is confining and, correspondingly, the
dual background does not have a horizon. In this case the wrapped
D5-brane again falls to the minimum value of the radial coordinate, but
its tension there is non-vanishing. This is the BPS domain wall
which separates adjacent inequivalent vacua distinguished by the
phase of the gluino condensate.

Finally, we show how to construct the
$SU(N)\times SU(N+M)$ theories mentioned above.
Consider
a D5-brane wrapped
over the 2-cycle, with its remaining directions
filling $\RR^{3,1}$.
If this object is located at some fixed $r$, then
it is a domain walls in $AdS_5$. The simplest domain wall is
a D3-brane which is not wrapped over the compact manifold.
Through an analysis of the five-form flux carried over
directly from \cite{Ed} one can conclude that when one crosses the
domain wall, the effect in field theory is to change the gauge group
from $SU(N) \times SU(N)$ to $SU(N+1) \times SU(N+1)$.

The field theory interpretation of a D5-brane wrapped around ${\bf S}^2$ is
more interesting:
if
on one side of the domain wall we have the original $SU(N) \times
SU(N)$ theory, then on the other side the theory is
$SU(N) \times SU(N+1)$ \cite{GK}.
The matter fields $A_k$ and $B_k$ are still
bifundamentals, filling out $2 (\N,\overline{\N+1}) \oplus 2
(\overline{\N}, \N+1)$.
One piece of evidence for this claim
is the way the D3-branes wrapped over the ${\bf S}^3$
behave when crossing the D5-brane domain wall.
In homology there is only one ${\bf S}^3$, but for
definiteness let us wrap the D3-brane around a particular three-sphere
${\bf S}^3_{(1)}$ which is invariant under the group $SU(2)_B$ under which
the fields $B_k$ transform.  The corresponding state in the $SU(N)
\times SU(N)$ field theory is ${\cal B}_1$ of (\ref{BaryonTwo}).  In the
$SU(N) \times SU(N+1)$ theory, one has instead
\[ 
   \epsilon_{\alpha_1 \ldots \alpha_N} \epsilon^{\beta_1 \ldots \beta_{N+1}}
    A^{\alpha_1}_{\beta_1} \ldots A^{\alpha_N}_{\beta_N}
      \qquad \hbox{or} \qquad
\]
\be
\qquad \qquad
   \epsilon_{\alpha_1 \ldots \alpha_N} \epsilon^{\beta_1 \ldots \beta_{N+1}}
    A^{\alpha_1}_{\beta_1} \ldots A^{\alpha_N}_{\beta_N}
     A^{\alpha_{N+1}}_{\beta_{N+1}}
\label{TwoBaryons}
\ee
 where we have omitted $SU(2)$ indices.  Either the upper index
$\beta_{N+1}$, indicating a fundamental of $SU(N+1)$, or the upper
index $\alpha_{N+1}$, indicating a fundamental of $SU(N)$, is free.

How can this be in supergravity?  The answer is simple: the wrapped
D3-brane must have a string attached to it. Indeed, after a wrapped
D3-brane has passed through the wrapped D5-brane domain wall,
it emerges with a
string attached to it due to the string creation by crossing D-branes
which together span 8 dimensions \cite{bdg,dfk}.
Calculating the tension of a wrapped D5-brane as a function of $r$
shows that it scales as $r^4/L^2$.
Hence, the domain wall is not stable, but in fact wants to move
towards $r=0$. We will assume that the wrapped
D5-branes ``fall'' behind the horizon and are replaced by their
flux in the SUGRA background. This gives a well-defined
way of constructing the SUGRA duals of
the $SU(N)\times SU(N+M)$ gauge theories.

The D5-branes wrapped over 2-cycles are examples of
a more general phenomenon.
For many singular spaces $Y_6$ there are fractional D3-branes
which can exist only within the singularity \cite{gipol,doug,GK,KN}.
These fractional D3-branes are D5-branes wrapped over (collapsed)
2-cycles at the singularity.
In the case of the
conifold, the singularity is a point.  The addition of $M$ fractional
branes at the singular point changes the gauge group to $SU(N+M)\times
SU(N)$; the four chiral superfields remain, now in the representation
$({\bf N+M},{\bf\overline N})$ and its conjugate,
as does the superpotential \cite{GK,KN}.
The theory is no longer conformal.  Instead, the relative gauge
coupling $g_1^{-2}-g_2^{-2}$ runs logarithmically, as pointed out in
\cite{KN}, where the supergravity equations corresponding to this
situation were solved to leading order in $M/N$.  In \cite{KT} this
solution was completed to all orders; the conifold suffers logarithmic
warping, and the relative gauge coupling runs logarithmically at all
scales.  The D3-brane charge, i.e. the 5-form flux,
decreases logarithmically as well.
However, the logarithm in the solution is not cut off at small
radius; the D3-brane charge eventually
becomes negative and the metric becomes
singular.

In \cite{KT} it was conjectured that this solution corresponds to a
flow in which the gauge group factors repeatedly drop in size by $M$
units, until finally the gauge groups are perhaps $SU(2M)\times SU(M)$
or simply $SU(M)$.  It was further suggested that the strong dynamics
of this gauge theory would resolve the naked singularity in the
metric.    The flow is
in fact an infinite series of Seiberg duality transformations --- a
``duality cascade'' --- in which the number of colors repeatedly drops
by $M$ units \cite{KS}.
Once the number of colors in the smaller gauge group is
fewer than $M$, non-perturbative effects become essential.  We will
show that these gauge theories have an exact anomaly-free $\ZZ_{2M}$
R-symmetry, which is broken dynamically, as in pure \none\ Yang-Mills
theory, to $\ZZ_2$.  In the supergravity, this occurs through the
deformation of the conifold.  In
short, the resolution of the naked singularity found in \cite{KT}
occurs through the chiral symmetry breaking of the gauge theory. The
resulting space, {\it a warped deformed conifold}, is completely
nonsingular and without a horizon, leading to confinement \cite{KS}.


\section{The RG cascade}

The addition of $M$ fractional 3-branes (wrapped D5-branes) at the
singular point changes the gauge group to $SU(N+M)\times SU(N)$.
Let us consider the effect on the dual supergravity background of
adding $M$ wrapped D5-branes.  The D5-branes serve as sources of the
magnetic RR 3-form flux through the $\S^3$ of $T^{1,1}$. Therefore,
the supergravity dual of this field theory involves $M$ units of the
3-form flux, in addition to $N$ units of the 5-form flux:
\be \label{qcond}
{1\over 4\pi^2 \alpha'}\int_{\S^3} F_3 = M\ ,
\qquad\qquad {1\over (4 \pi^2 \alpha')^2 }\int_{T^{1,1}} F_5 = N
\ .
\ee
The coefficients above follow from the quantization rule
(\ref{quantten}).  The warped conifold solution with such fluxes was
constructed in \cite{KT}.

It will be useful to employ
the following basis of 1-forms on the compact space
\cite{MT}:
\bea \label{fbasis}
g^1 = {e^1-e^3\over\sqrt 2}\ ,\qquad
g^2 = {e^2-e^4\over\sqrt 2}\ , \nonumber \\
g^3 = {e^1+e^3\over\sqrt 2}\ ,\qquad
g^4 = {e^2+ e^4\over\sqrt 2}\ , \nonumber \\
g^5 = e^5\ ,
\eea
where
\begin{eqnarray}
e^1\equiv - \sin\theta_1 d\phi_1 \ ,\qquad
e^2\equiv d\theta_1\ , \nonumber \\
e^3\equiv \cos\psi\sin\theta_2 d\phi_2-\sin\psi d\theta_2\ , \nonumber\\
e^4\equiv \sin\psi\sin\theta_2 d\phi_2+\cos\psi d\theta_2\ , \nonumber \\
e^5\equiv d\psi + \cos\theta_1 d\phi_1+ \cos\theta_2 d\phi_2 \ .
\end{eqnarray}
In terms of this basis, the Einstein metric on $T^{1,1}$ assumes the
form
\be
ds^2_{T^{1,1}}= {1\over 9} (g^5)^2 + {1\over 6}\sum_{i=1}^4 (g^i)^2
\ .
\ee

Keeping track of the normalization factors, in order
to be consistent with the quantization conditions (\ref{qcond}),
\be \label{closedf}
F_3 = {M\alpha'\over 2}\omega_3\ ,  \qquad\qquad
B_2 = {3 g_s M \alpha'\over 2}\omega_2 \ln (r/r_0)
\ ,
\ee
\be
H_3 = dB_2 = {3 g_s M \alpha' \over 2r} dr\wedge \omega_2\ ,
\label{hthree}
\ee
where
\be
\omega_2 = {1\over 2}(g^1\wedge g^2 + g^3 \wedge g^4)=
{1\over 2} (\sin\theta_1 d\theta_1 \wedge d\phi_1-
\sin\theta_2 d\theta_2 \wedge d\phi_2 )\ ,
\label{omegato}
\ee
\be
\omega_3 = {1\over 2} g^5\wedge (g^1\wedge g^2 + g^3 \wedge g^4)\ .
\label{omegathr}
\ee
One can show that \cite{remarks}
\be
\int_{\S^2} \omega_2 = 4 \pi\ , \qquad
\int_{\S^3} \omega_3 = 8 \pi^2 \
\label{intforms}
\ee
where the $S^2$ is parametrized by $\psi=0$, $\theta_1=\theta_2$ and
$\phi_1=-\phi_2$, and the $S^3$ by $\theta_2=\phi_2=0$.  As a result,
the quantization condition for RR 3-form flux is obeyed.

Both $\omega_2$ and $\omega_3$ are closed.
Note also that
\be \label{duality}
g_s \star_6 F_3 = H_3\ ,\qquad g_s F_3 =  -\star_6 H_3\ ,
\ee
where $\star_6$ is the Hodge dual with respect to the metric
$ds_6^2$. Thus, the complex 3-form $G_3$
satisfies the self-duality condition
\be
\star_6 G_3 = i G_3\ , \qquad\qquad G_3 = F_3 - {i\over g_s} H_3\ .
\ee
Note that the self-duality fixes the relative factor of 3 in
(\ref{closedf}) (see (\ref{conimetric}), (\ref{co})). We will see that
this geometrical factor is crucial for reproducing the well-known
factor of 3 in the ${\cal N}=1$ beta functions.

It follows from (\ref{duality}) that
\be \label{dilcons}
g_s^2 F_3^2 = H_3^2
\ ,
\ee
which implies that the dilaton is constant, $\Phi=0$.
Since $F_{3\mu\nu\lambda} H_3^{\mu\nu\lambda} =0$, the RR scalar
vanishes as well.

The 10-d metric found in \cite{KT} has the structure of a
``warped product'' of $\R{3,1}$ and the conifold:
\be \label{fulsol}
ds^2_{10} =   h^{-1/2}(r)   dx_n dx_n
 +  h^{1/2} (r)  (dr^2 + r^2 ds^2_{T^{1,1}} )\ .
\ee
The solution
for the warp factor $h$ may be determined
from the trace of the Einstein equation:
\be
R = {1\over 24} (H_3^2 + g_s^2 F_3^2) = {1\over 12} H_3^2\ .
\ee
This implies
\be
- h^{-3/2} {1\over r^5} {d\over dr} (r^5 h') ={1\over 6} H_3^2
\ .
\ee
Integrating this differential equation, we find that
\be \label{nonharm}
h(r) = {27 \pi (\alpha')^2 [g_s N +
a (g_s M)^2 \ln (r/r_0) + a (g_s M)^2/4]
\over 4 r^4}
\ee
with $a=3/(2\pi)$.

An important feature of this background
is that $\tilde F_5$ acquires a radial dependence \cite{KT}.
This is because
\be
\tilde F_5 = F_5 + B_2\wedge F_3\ , \qquad F_5 = d C_4\ ,
\ee
and $\omega_2\wedge \omega_3 = 54 {\rm vol}(T^{1,1})$.
Thus, we may write
\be
\tilde F_5 = {\cal F}_5 + \star {\cal F}_5 \ , \qquad  {\cal F}_5
= 27 \pi \alpha'^2 N_{eff} (r) {\rm vol}({\rm T}^{1,1})
\ ,
\label{F5KT}
\ee
and
\be
N_{eff} (r) = N + {3\over 2 \pi} g_s M^2 \ln (r/r_0)
\ .
\label{Neff}
\ee
The novel phenomenon in this solution is that the 5-form flux present
at the UV scale $r=r_0$ may completely disappear by the time we
reach a scale where $N_{eff} = 0$.
The non-conservation of the flux is due to the type IIB SUGRA equation
\be \label{transg}
d \tilde F_5 = H_3 \wedge F_3
\ .
\ee
A related fact is that
$\int_{\S^2} B_2$ is
no longer a periodic variable in the SUGRA solution
once the $M$ fractional branes are introduced:
as the $B_2$ flux goes through a period, $N_{eff} (r) \rightarrow
N_{eff} (r) - M$
which has the effect of decreasing the 5-form flux by $M$ units.
Note from (\ref{Neff}) that for a single cascade step
$N_{eff} (r) \rightarrow N_{eff} (r) - M$
the radius changes by a factor
$r_2 / r_1 = \exp(-2\pi / 3 g_s M)$, agreeing with a result of
\cite{GKP}.

Due to the non-vanishing RHS of (\ref{transg}),
${1\over (4 \pi^2 \alpha')^2 }\int_{T^{1,1}} \tilde F_5$
is not quantized. We may identify this quantity with $N_{eff}$
defining the gauge group $SU(N_{eff}+M)\times SU(N_{eff})$
only at special radii $r_k= r_0 \exp(-2\pi k/ 3 g_s M)$ where
$k$ is an integer. Thus, $N_{eff}= N - kM$.
Furthermore, we believe that the continuous
logarithmic variation of $N_{eff} (r)$
is related to continuous reduction in the number of degrees of
freedom as the theory flows to the IR. Some support for this claim
comes from studying the high-temperature phase of this theory
using black holes embedded into asymptotic KT geometry
\cite{finitetemp}. The effective number of degrees of freedom
computed from the Bekenstein--Hawking entropy grows logarithmically with
the temperature, in agreement with (\ref{Neff}).

The metric (\ref{fulsol})
has a naked singularity at $r=r_s$ where $h(r_s)=0$.
Writing
\be \label{UVs}
h(r) = {L^4\over r^4} \ln (r/r_s)\ , \qquad
L^2 = {9  g_s M \alpha'\over 2\sqrt 2}
\ ,
\ee
we find a purely logarithmic RG cascade:
\be \label{dualmet}
ds^2 = {r^2\over L^2 \sqrt{\ln (r/r_s)} } dx_n dx_n
+ { L^2 \sqrt{\ln (r/r_s)}\over r^2} dr^2 + L^2 \sqrt{\ln (r/r_s)}
ds^2_{T^{1,1}}\ .
\ee
Since $T^{1,1}$ expands
slowly toward large $r$, the curvatures decrease there so
that corrections to the SUGRA become negligible. Therefore,
even if $g_s M$ is very small,
this SUGRA solution is reliable for sufficiently
large radii where $g_s N_{eff} (r)\gg 1$.
In this regime the separation
between the cascade steps is very large, so that the SUGRA
calculation of the $\beta$-functions may be compared with
$SU(N_{eff}+ M)\times SU(N_{eff})$ gauge theory.
We will work near $r=r_0$ where $N_{eff}$ may be replaced by $N$.

\subsection{Matching of the $\beta$-functions}

In order to match the two
gauge couplings to the moduli of the type IIB
theory on $AdS_5\times T^{1,1}$, one notes that the integrals over the
$\S^2$ of $T^{1,1}$ of the NS-NS and R-R 2-form potentials, $B_2$ and
$C_2$, are moduli. In particular, the two gauge couplings are
determined as follows \cite{KW,MP}:\footnote{Exactly the same relations
apply to the ${\cal N}=2$ supersymmetric $\ZZ_2$ orbifold theory
\cite{ks,Joe}.}
\be
{4\pi^2 \over g_1^2} + {4\pi^2 \over g_2^2} ={\pi\over g_s e^\Phi}
\ ,
\ee
\be \label{couplediff}
\left [ {4\pi^2 \over g_1^2} - {4\pi^2 \over g_2^2}\right ]
g_s e^\Phi
= {1\over 2\pi \alpha'}
\left(\int_{\S^2} B_2\right) - \pi \quad ({\rm mod}\ 2\pi)
\ .
\ee
From
the quantization condition on $H_3$,
${1\over 2\pi \alpha'}(\int_{\S^2} B_2)$
must be a
periodic variable with period $2\pi$.
This periodicity is crucial for the cascade phenomenon.
These equations are crucial for relating the SUGRA
background to the field theory $\beta$-functions when the
theory is generalized to $SU(N+M)\times SU(N)$ \cite{KN,KT}.

In gauge/gravity duality the 5-dimensional radial coordinate
defines the RG scale of the dual gauge theory
\cite{jthroat,US,EW,holobound,uvir}.
There are different ways of establishiing the precise relation.
The simplest one is to identify the field theory energy scale
$\Lambda$
with the energy of a stretched string ending on a probe brane
positioned at radius $r$. For all metrics of the form (\ref{fulsol})
this gives
\be
\Lambda\sim r
\ .
\ee
In this section we adopt this UV/IR relation, which typically
corresponds to the Wilsonian renormalization group.

Now we are ready to interpret the solution of \cite{KT} in terms
of RG flow in the dual $SU(N+M)\times SU(N)$ gauge theory.
The constancy of the dilaton translates into the vanishing
of the $\beta$-function for
$ {8\pi^2\over g_1^2} + {8\pi^2\over g_2^2}$.
Substituting the solution for $B_2$ into (\ref{couplediff})
we find
\be \label{betres}
{8\pi^2\over g_1^2} - {8\pi^2\over g_2^2} = 6 M \ln (r/r_s) + {\rm
const}
\ .
\ee
Since $\ln (r/r_s) = \ln (\Lambda/\mu)$,  (\ref{betres})
implies a logarithmic running of
${1\over g_1^2} - {1\over g_2^2}$ in the $SU(N+M)\times SU(N)$ gauge theory.
As we mentioned earlier, this SUGRA result is reliable
for any value of $g_s M$ provided that
$g_s N \gg 1$. We may consider, for instance,
$g_s M \ll 1$ so that the cascade
jumps are well-separated.

Let us compare with the Shifman--Vainshtein
$\beta$-functions \cite{NSVZ}:\footnote{
These expressions for the $\beta$-functions
differ from the standard NSVZ form
\cite{NSVZinstanton} by a factor of $1/(1 - g^2 N_c / 8\pi^2)$.
The difference comes from the choice of
normalization of the vector superfields.
We choose the normalization so that the relevant kinetic term in the
field theory action is $\frac{1}{2g^2}\int d^4x d^2\theta
Tr(W^{\alpha} W_{\alpha})+$ h.c.;
this choice is dictated by the form of the supergravity action and
differs from the canonical normalization by a factor of $1/g^2$.
With this convention the additional factor in the $\beta$-function does
not appear.
A nice review of the derivation of the exact
$\beta$-functions is in \cite{susynotes}.}
\beq \label{SVexact}
&{d\over d {\rm log} (\Lambda/\mu)}
{8 \pi^2\over g_1^2} & = 3(N +M) - 2N (1- \gamma)\ ,\\
&{d\over d {\rm log} (\Lambda/\mu)}
{8 \pi^2 \over g_2^2} & = 3N - 2(N+M) (1-  \gamma)\ ,
\eeq
where $\gamma$ is the anomalous dimension
of operators ${\rm Tr} A_i B_j$.
The conformal invariance of the field theory for $M=0$,
and symmetry under $M\rightarrow -M$,
require that
$\gamma =-{1\over 2} + O[(M/N)^{2n}]$ where $n$
is a positive integer \cite{KS}.
Taking the difference of the two equations in (\ref{SVexact}) we then
find
\begin{eqnarray} \label{betafun}
{8\pi^2 \over g_1^2} - {8\pi^2 \over g_2^2} &=&
 M \ln (\Lambda/\mu) [ 3 + 2(1-\gamma)] \\ 
&=& 6 M \ln (\Lambda/\mu) (1+ O[(M/N)^{2n}])\ .
\nonumber
\end{eqnarray}
Remarkably, the coefficient $6M$ is in {\it exact} agreement
with the result
(\ref{betres}) found on the SUGRA side.
This consitutes a geometrical explanation of a field theory $\beta$-function,
including its normalization.

We may also
trace the jumps in the rank of the gauge group to a well-known
phenomenon in the dual ${\cal N}=1$ field theory, namely, Seiberg
duality \cite{NAD}. The essential observation is that $1/g_1^2$ and
$1/g_2^2$ flow in opposite directions and, according to
(\ref{SVexact}), there is a scale where the $SU(N+M)$ coupling, $g_1$,
diverges. To continue past this infinite coupling, we perform a ${\cal
N}=1$ duality transformation on this gauge group factor.  The
$SU(N+M)$ gauge factor has $2N$ flavors in the fundamental
representation.  Under a Seiberg duality transformation, this becomes
an $SU(2N-[N+M]) = SU(N-M)$ gauge group.
Thus we obtain an $SU(N)\times SU(N-M)$
theory which resembles closely the theory we started with \cite{KS}.

As the theory flows to the IR, the cascade must stop, however, because
negative $N$ is physically nonsensical. Thus, we should not
be able to continue the solution (\ref{dualmet}) to
the region where $N_{eff}$ is negative.
To summarize,
the fact that the solution of \cite{KT} is singular
tells us that it has to be modified in the IR. The necessary modification
proceeds via the deformation of the conifold, and is discussed in section
5.

\section{The Chiral Anomaly}

In theories with ${\cal N}=1$ supersymmetry, $\beta$-functions
are related to chiral anomalies \cite{NSVZ}. The essential
mechanism is the $\beta$-functions contribute to the trace
anomaly, $\langle T^i_i\rangle$, which is related by
supersymmetry to the divergence of the $U(1)_R$
current, $\partial_i J^i$. In the previous section we showed how
the logarithmic running of the gauge couplings manifests itself
in the dual supergravity solution of \cite{KT}.
Here we show that the chiral anomaly
can be read off the solution as well.
Although the metric has a continuous
$U(1)_R$ symmetry, the full supergravity solution is only
invariant under a $\ZZ_{2M}$
subgroup of this $U(1)$.  In the dual quantum field theory there are
chiral fermions charged under the $U(1)_R$, and so we can understand
the R-symmetry breaking as an effect of the chiral anomaly.  Anomalies
are especially interesting creatures for the gauge/gravity duality,
because the Adler-Bardeen theorem \cite{adler} guarantees that anomaly
coefficients computed at one loop are exact, with no radiative corrections; the
significance of this fact is that we can compute anomaly coefficients
in the field theory at weak coupling, then extrapolate the results to
strong coupling, where we can use dual gravity methods to check the
calculation.  In this section we will study some aspects of the
anomaly in detail for the cascading gauge theory.

There are three lessons that we can take away from this
analysis \cite{kow}. First, the anomaly coefficients computed on each
side of the duality agree exactly, even for our non-conformal
cascading theory with only ${\cal N} = 1$ supersymmetry; although this
result is hardly surprising, it is a nice check of the duality.
Second, the symmetry breaking is a classical effect on the gravity
side.  There is no need to appeal to instantons, which is a good thing
as they do not appear anywhere explicitly in the gravity dual.
Finally, the R-symmetry is broken {\it spontaneously} in the
supergravity solution -- the bulk vector field dual to the R-current
of the gauge theory acquires a mass.  The symmetry breaking then
appears ``anomalous'' if one insists on a four-dimensional
description.

\subsection{The Anomaly as a Classical Effect in Supergravity}

The asymptotic UV metric (\ref{fulsol},\ref{co}) has
a $U(1)$ symmetry associated with the rotations of the angular
coordinate $\beta= \psi/2$,
normalized so that $\beta$ has period $2\pi$.
This is the R-symmetry of the dual gauge
theory.  It is crucial, however, that the background value of the R-R
2-form $C_2$ does not have this continuous symmetry.  Indeed, although
$F_3$ is $U(1)$ symmetric, there is no smooth global expression for
$C_2$. Locally, we may write 
\be \label{orig} C_2\rightarrow M\alpha'\beta \omega_2 \ .\ee
This expression is not single-valued as a function of the angular
variable $\beta$, but it is single-valued up to a gauge
transformation, so that $F_3=dC_2$ is single-valued.  In fact, $F_3$
is completely independent of $\beta$.  Because of the explicit $\beta$
dependence, $C_2$ is not $U(1)$-invariant. Under the
transformation $\beta\rightarrow \beta+\epsilon$,
\be \label{shift}
C_2 \rightarrow C_2 + M\alpha' \epsilon \omega_2
\ .\ee
Since $\int_{\S^2} C_2$ is defined modulo 
$4\pi^2 \alpha'$,
a gauge transformation can shift $C_2/(4\pi^2\alpha')$ by an arbitrary
integer multiple of $\omega_2/(4\pi)$, so $\beta\to\beta+\epsilon$ is
a symmetry precisely if $\epsilon$ is an integer multiple of
$\pi/M$. 
Because $\epsilon$ is anyway only defined mod $2\pi$, a
$\ZZ_{2M}$ subgroup of the $U(1)$ leaves fixed the asymptotic values
of the fields, and thus corresponds to a symmetry of the system.
This $\ZZ_{2M}$ is a symmetry since it
respects the asymptotic values of the fields.

Let us compare the above analysis with the gauge theory. 
(A similar comparison for
the case of an ${\cal N} = 2$ orbifold theory appeared in \cite{bertd7}.)
As pointed out
in \cite{KW}, the integral of the RR 2-form potential $C_2$ over the
${\bf S}^2$ of $T^{1,1}$ is a modulus.  Because the integral of $B_2$
was dual to the difference of gauge couplings for the two gauge
groups, it is natural that the integral of $C_2$ is dual to
the difference of $\Theta$-angles (it is possible to check this
statement explicitly in orbifold backgrounds).
The $\Theta$-angles are given by
\be \label{diffsum} \Theta_1 - \Theta_2 ={1\over \pi\alpha'}
\int_{S^2} C_2 \ ,\qquad \Theta_1 + \Theta_2 \sim C \ , \ee
where $C$ is the RR scalar, which vanishes for the case under
consideration.
Using the fact that
$\int_{\S^2} \omega_2 = 4\pi$, we find that the small $U(1)$ rotation
$\beta \rightarrow \beta+ \epsilon$ induces
\be \label{angles}
\Theta_1 =-\Theta_2 = 2 M\epsilon
\ .
\ee
With a conventional normalization, the $\Theta$ terms appear in the gauge
theory action as
\be
\int d^4 x ({\Theta_1\over 32 \pi^2}
 F^a_{ij} \tilde F^{aij} +  {\Theta_2\over 32 \pi^2}
G^b_{ij} \tilde G^{bij})\ .
\label{thetaaction}
\ee
If we assume that $\epsilon$ is a function of the 4 world volume
coordinates $x^i$, then under the $U(1)$ rotation (\ref{angles})
the terms linear in $\epsilon$ in the dual
gauge theory (\ref{thetaaction}) are
\be \label{holanom}
\int d^4 x [- \epsilon \partial_i J^i +
{M\epsilon \over 16 \pi^2} (
 F^a_{ij} \tilde F^{aij} -  G^b_{ij} \tilde G^{bij} )
]\ ,
\ee
where $J^i$ is the chiral $R$-current.  The appearance of the second
term is due to the non-invariance of $C_2$ under the $U(1)$ rotation.
Varying with respect to $\epsilon$, we therefore obtain
\be
 \partial_i J^i = {M \over 16 \pi^2}
\left(
 F^a_{ij} \tilde F^{aij} -  G^b_{ij} \tilde G^{bij} \right) \ .
\label{anomeq}
\ee

This anomaly equation, derived from supergravity, agrees exactly
with our expectations from the gauge theory.
A standard result of quantum field theory is that in a theory with
chiral fermions charged under a global $U(1)$ symmetry of the classical
Lagrangian, the Noether current associated with that symmetry is not
generally conserved but instead obeys the equation
\beq
\partial_i J^i =  \frac{1}{32\pi^2} \sum_m n_m R_m  F^a_{ij} \tilde F^{aij}
\eeq
where $n_m$ is the number of chiral fermions with R-charge $R_m$
circulating in the loop of the relevant triangle diagram.  In the case
of interest, there are two gauge groups, so let us define $F^a_{ij}$
and $G^b_{ij}$ to be the field strengths of $SU(N+M)$ and $SU(N)$
respectively.  Now, the chiral superfields $A_i, B_j$ contribute $2N$ flavors
to the gauge group $SU(N+M)$, and each one carries R-charge $1/2$. The
chiral fermions which are their superpartners have R-charge $-1/2$
while the gluinos have R-charge $1$. Therefore, the anomaly
coefficient is ${M \over 16 \pi^2}$.  An equivalent calculation for
the $SU(N)$ gauge group with $2(N+M)$ flavors produces the opposite
anomaly, so the anomaly equation as computed from field theory is
just (\ref{anomeq}).


The upshot of the calculation presented above is that the chiral
anomaly of the $SU(N+M)\times SU(N)$ gauge theory is encoded in
the ultraviolet (large $r$) behavior of the dual classical
supergravity solution. No additional fractional D-instanton
effects are needed to explain the anomaly. Thus, as often occurs
in the gauge/gravity duality, a quantum effect on the gauge theory
side turns into a classical effect in supergravity.
Similar methods have been used to describe chiral anomalies
in other supersymmetric gauge theories \cite{kow,bertd7,mbert}.

\subsection{The Anomaly as Spontaneous Symmetry Breaking in $AdS_5$}

Let us look for a deeper understanding of the anomaly from the dual
gravity point of view.  On the gauge theory side, the R-symmetry is
global, but in the gravity dual it as usual becomes a gauge symmetry,
which must not be anomalous, or the theory would not make sense at
all.  Rather, we will find that the gauge symmetry is spontaneously
broken: the 5-d vector field dual to the R-current of the gauge theory
`eats' the scalar dual to the difference of the theta angles and
acquires a mass.\footnote{
The connection between anomalies in a D-brane field theory
and spontaneous symmetry breaking in string theory was previously
noted in \cite{aks} (and probably elsewhere in the literature).}
A closely related mechanism was observed in studies of RG flows from
the dual gravity point of view \cite{BDFP,BFS}. There R-current
conservation was violated not through anomalies but by turning on
relevant perturbations or expectation values for fields. In these
cases it was shown \cite{BDFP,BFS} that the 5-d vector field dual to
the R-current acquires a mass through the Higgs mechanism. We will
show that symmetry breaking through anomalies can also have the bulk
Higgs mechanism as its dual.

In the absence of fractional branes there are no background three-form
fluxes, so the $U(1)$ R-symmetry is a true symmetry of the field
theory.  Because the R-symmetry is realized geometrically by
invariance under a rigid shift of the angle $\beta$, it becomes a
local symmetry in the full gravity theory, and the associated gauge
fields $A = A_{\mu} dx^{\mu}$ appear as fluctuations of the
ten-dimensional metric and RR four-form potential \cite{Kim,Ceres}.
The natural metric ansatz is of the familiar Kaluza-Klein form:
\begin{eqnarray}
  ds^2 &=& h(r)^{-1/2} \left( dx_n dx^n \right) +
\nonumber \\
 && h(r)^{1/2} r^2 \left[\frac{dr^2}{r^2} +
\frac{1}{9} \left(g^5 - 2A \right)^2+
          \frac{1}{6} \sum_{r=1}^4 \left( g^{r}\right)^2 \right],
\label{kkmetric}
\end{eqnarray}
where $h(r) = L^4/r^4$, and $L^4 = \frac{27}{16}(4\pi \alpha'^2 g_s
N)$.  It is convenient to define the one-form $\chi = g^5 - 2A$, which
is invariant under the combined gauge transformations
\beq
\beta \rightarrow \beta + \lambda,
\qquad A \rightarrow A + d\lambda.
\label{gaugetrans}
\eeq
The equations of motion for the field $A_{\mu}$ appear as the $\chi\mu$
components of Einstein's equations,
\beq
   R_{MN} = \frac{g_s^2}{4 \cdot 4!} \tilde{F}_{MPQRS} \tilde{F}_N^{~PQRS}.
\eeq

The five-form flux will also fluctuate when we activate the
Kaluza-Klein gauge field; indeed, the unperturbed $\tilde{F}_5$ of
(\ref{F5KT}) is not self-dual with respect to the gauged metric
(\ref{kkmetric}).  An appropriate ansatz to linear order in $A$ is
\beq \label{fiveform}
   \tilde{F}_5= dC_4 &=& \frac{1}{g_s} d^4 x \wedge dh^{-1} +
   \frac{\pi \alpha'^2 N}{4} \Bigg[\chi \wedge
   g^1 \wedge g^2 \wedge g^3 \wedge g^4 \nonumber \\
  & & \qquad \qquad - dA \wedge g^5 \wedge dg^5 +
  \frac{3}{L} \star_5 dA \wedge dg^5 \Bigg].
\eeq
The five-dimensional Hodge dual $\star_5$ is defined with respect to
the AdS$_5$ metric $ds_5^{~2}=h^{-1/2} dx_n dx^n + h^{1/2} dr^2$.  It
is straightforward to show that the supergravity field equation $
d\tilde F_5 =0$ implies that the field $A$ satisfies the equation of
motion for a massless vector field in AdS$_5$ space:
\beq d \star_5 dA =0 \ .
\label{vecteq}
\eeq
Using the identity $dg^5 \wedge dg^5 = -2 g^1 \wedge g^2 \wedge g^3
\wedge g^4$, we can check that the expression for $C_4$ is
%
\beq
C_4 &=& {1\over g_s} h^{-1} d^4 x + \frac{\pi \alpha'^2 N}{2}
\Bigg[ \beta g^1 \wedge g^2 \wedge g^3 \wedge g^4 - {1\over 2}
A\wedge d g^5 \wedge g^5 \nonumber \\
& & \qquad \qquad \qquad -\frac{3}{2r} h^{-1/4} \star_5 dA \wedge g^5 \Bigg].
\label{cfourads}
\eeq
Another way to see that $A$ is a massless vector in $AdS_5$ is to
consider the Ricci scalar for the metric (\ref{kkmetric})
\beq
  R = R(A=0) - \frac{h^{1/2}r^2}{9} F_{\mu \nu} F^{\mu \nu}
\label{riccikk}
\eeq
so that on reduction from ten dimensions the five-dimensional
supergravity action will contain the action for a massless vector
field.

The story changes when we add wrapped D5-branes.  As described in
Section 2, the 5-branes introduce $M$ units of RR flux through the
three-cycle of $T^{1,1}$.  
Now, the new wrinkle is that the RR three-form flux
of (\ref{closedf}) is not gauge-invariant with respect to shifts of
$\beta$ (\ref{gaugetrans}).  To restore the gauge invariance, we
introduce a new field $\theta\sim \int_{S^2} C_2$:
\beq
F_3 = dC_2 = \frac{M\alpha'}{2} \left( g^5 +
              2 \partial_{\mu} \theta dx^{\mu} \right) \wedge \omega_2
\eeq
so that $F_3$ is invariant under the gauge transformation $\beta
\rightarrow \beta + \lambda$, $\theta \rightarrow \theta -\lambda$.
Let us also define $W_{\mu} = A_{\mu} + \partial_{\mu} \theta$.  In
terms of the gauge invariant forms $\chi$ and $W=W_{\mu}dx^{\mu}$,
\beq
F_3 = \frac{M\alpha'}{2} (\chi + 2W) \wedge \omega_2.
\label{f3shift}
\eeq

{}From (\ref{f3shift}) we can immediately see how the anomaly will
appear in the gravity dual.  Assuming that the NS-NS three form is
still given by (\ref{hthree}), we find that up to terms of order
$g_sM^2/N$ the three-form equation implies
\beq
d\star_5 W = 0 \, \Rightarrow \, \frac{L^2}{r^2} \partial_i W^i +
\frac{1}{r^5} \partial_r r^5 W_r = 0
\label{dstarw}
\eeq
which is just what one would expect for a massive vector field in five
dimensions.  To a four dimensional observer, however, a massive vector
field would satisfy $\partial_i W^i=0$.  Thus in the field theory one
cannot interpret the $U(1)$ symmetry breaking as being spontaneous,
and the additional $W_r$ term in (\ref{dstarw}) appears in four
dimensions to be an anomaly.

Another way to see that the vector field becomes massive is to compute
its equation of motion.  To do this calculation precisely, we should
derive the $\chi \mu$ components of Einstein's equations, and also
find the appropriate expressions for the five-form and metric up to
quadratic order in $g_s M$ and linear order in fluctuations.  This
approach is somewhat nontrivial.  A more heuristic 
approach is to consider the type IIB supergravity action
to quadratic order in $W$, ignoring the 5-form field strength
contributions:
\beq
   S & = & -\frac{1}{2\kappa_{10}^{~2}} \int d^{10}x \sqrt{-G_{10}}
\left[ R_{10} -\frac{g_s^{~2}}{12}|F_3|^2 \right] + \ldots \\
    &\sim& -\frac{1}{2\kappa_{10}^{~2}} \int d^{10}x \sqrt{-G_{10}}
    \left[ - \frac{h^{1/2}r^2}{9} F_{\mu \nu} F^{\mu \nu} - \right.
\nonumber \\
&& \qquad \qquad \qquad
    \left. \left(\frac{g_s M\alpha'}{2} \right)^2
     \frac{36}{hr^4} W_{\mu} W^{\mu} \right] + \ldots
\label{action}
\eeq
This is clearly the action for a massive four-dimensional vector
field, which has as its equation of motion
\beq
\partial_{\mu} (hr^7 F^{\mu \nu}) = \tilde m^2 hr^7 W^{\nu}
\label{weom}
\eeq
which in differential form notation is $d(h^{7/4}r^7\star_5 dW) = -
\tilde m^2 h^{7/4}r^7 \star_5 W$.  {}From the action (\ref{action}),
we see that the mass-squared is given by
\beq
 \tilde m^2 &=& \left(g_s M\alpha' \right)^2 \frac{81}{2h^{3/2} r^6}.
\label{mass}
\eeq
%
%
This result, however, ignores the subtlety of the type IIB action in
presence of the self-dual 5-form field. 
A more precise calculation \cite{krasnitz}, which takes the mixing into account,
gives instead the following equation for the transverse vector modes:
\beq \label{mk}
\left ({1 \over hr^7}\partial_r hr^7 \partial_r +
h\partial_{i}\partial_{i}-{(9 M\alpha')^4 \over 64 h^2 r^{10} }\right )W_{i}=0,
\eeq
This shows that the 10-d mass actually appears at a higher order in
perturbation theory compared to the result (\ref{mass})
that ignores the mixing
with the 5-form.

Let us compare this result to earlier work.
In \cite{BDFP,brand,BFS} it was shown that the
5-d vector field associated with a $U(1)_R$ symmetry acquires a mass
in the presence of a symmetry-breaking relevant perturbation, and that
this mass is related in a simple way to the warp factor of the
geometry.\footnote{We are grateful to O.~DeWolfe and K.~Skenderis for
pointing out the relevance of this work to the present calculation.}
It is conventional to write the 5-d gauged supergravity metric in the
form
\beq
\tilde{G}_{\mu \nu} dx^{\mu}dx^{\nu} = e^{2T(q)}\eta_{ij} dx^i dx^j +dq^2.
\label{dwmetric}
\eeq
The result of \cite{BDFP} is that $m^2 = -2T''$.  To relate
the 5-d metric (\ref{dwmetric}) to the 10-d metric (\ref{specans})
we must normalize the 5-d metric so that the graviton
has a canonical kinetic term.  Doing this carefully we find
\beq
\tilde{G}_{\mu \nu}dx^{\mu}dx^{\nu} =
\left( hr^4/L^4 \right)^{5/6} (h^{-1/2}\eta_{ij} dx^i dx^j+h^{1/2}dr^2).
\label{metric5d}
\eeq
The factor $\left( hr^4/L^4 \right)^{5/6}$ arises due to the
radial dependence of the size of $T^{1,1}$ through the
usual Kaluza--Klein reduction.
The radial variables $q$ and $r$ are related,
at leading order in $g_s M^2/N$, by
\beq
\log(r) \sim \frac{q}{L} - \frac{g_sM^2}{2\pi N}\left(\frac{q}{L}\right)^2.
\eeq
We can also show that $-2T = -2\log(r) +$(terms which do not affect
the mass to leading order in $g_s M^2/N$), so now computing the
mass-squared by the prescription of \cite{BDFP} we 
obtain
\beq
  m^2 = \frac{4}{\alpha' (3\pi)^{3/2} } \frac{(g_s M)^2}{(g_s N)^{3/2}}.
\label{rightmass}
\eeq
where this mass applies to a vector field $V$ with a canonical
kinetic term for the metric (\ref{metric5d}).
For these calculations it is convenient to work with the transverse
4-d vector modes $V_i$ and to decouple the longitudinal modes such as $V_r$.
The equation of motion of $V$ is
\beq
(e^{-2T} \frac{\partial}{\partial q}e^{2T} \frac{\partial}{\partial q} +
e^{-2T}\partial_i \partial_i - m^2)V_i = 0.
\eeq
In fact, this equation follows from (\ref{mk}) after a rescaling
\cite{krasnitz}
\beq
V_i = (hr^4/L^4)^{2/3} W_i.
\eeq
The nonvanishing vector mass is consistent with gauge invariance because
the massless vector field $A$ has eaten the scalar field $\theta$,
spontaneously breaking the gauge symmetry, as advertised.  It is
interesting that the anomaly appears as a bulk effect in
AdS space, in contrast to some earlier examples
\cite{EW,henningson} where anomalies arose from boundary terms.

The appearance of a mass implies that the R-current operator should
acquire an anomalous dimension.
{}From (\ref{rightmass}) it follows that
\be (m L)^2 = \frac{2 (g_s M)^2}{\pi (g_s N)}\ .
\ee
Using the AdS/CFT correspondence (perhaps naively, as the KT metric is not asymptotically AdS)
 we find that the dimension of the
current $J^\mu$ dual to the vector field $W^\mu$ is
\be
\Delta = 2 + \sqrt{1 + (mL)^2} \ .
\ee
Therefore, the anomalous dimension of the current is
\be
\Delta - 3 \approx (mL)^2/2 =\frac{(g_s M)^2}{\pi (g_s N)}\ .
\label{anomdim}
\ee
We can obtain a rough understanding of this result by considering the
relevant weak coupling calculation in the gauge theory.  The leading
correction to the current-current two-point function comes from the
three-loop Feynman diagram composed of two triangle diagrams glued
together, and the resulting anomalous dimension $\gamma_J$ is
quadratic in $M$ and $N$.  $\gamma_J$ must vanish when $M=0$, and it
must be invariant under the map $M \rightarrow -M$, $N \rightarrow
N+M$, which simply interchanges the two gauge groups.  Thus, the
lowest order piece of the anomalous dimension will be of order $(g_s
M)^2$.  Our supergravity calculation predicts that this anomalous
dimension is corrected at large $g_s N$ by an extra factor of $1/(g_s
N)$.  Of course, it would be interesting to understand this result
better from the gauge theory point of view.

\section{Deformation of the Conifold}

It was shown in \cite{KS}
that, to remove the naked singularity found in
\cite{KT} the conifold (\ref{coni}) should be replaced by the deformed
conifold
\begin{equation} \label{dconifold}
\sum_{i=1}^4 z_i^2 =
\varepsilon^2\ ,
\end{equation}
in which the singularity of the conifold is removed
through the blowing-up of the  $\S^3$ of $T^{1,1}$.
The 10-d metric of \cite{KS}
takes the following form:
\be \label{specans}
ds^2_{10} =   h^{-1/2}(\tau)   dx_n dx_n
 +  h^{1/2}(\tau) ds_6^2 \ ,
\ee
where $ds_6^2$ is the metric of the deformed conifold (\ref{metricd}).
This is the same type of ``D-brane'' ansatz as (\ref{fulsol}), but with the
conifold replaced by the deformed conifold as
the transverse space.

The metric of the deformed conifold was discussed in some detail in
\cite{cd,MT,Ohta}. It is diagonal in the basis (\ref{fbasis}):
\bea \label{metricd}
ds_6^2 = {1\over 2}\varepsilon^{4/3} K(\tau)
\Bigg[ {1\over 3 K^3(\tau)} (d\tau^2 + (g^5)^2)
 +
\cosh^2 \left({\tau\over 2}\right) [(g^3)^2 + (g^4)^2]\nonumber \\
+ \sinh^2 \left({\tau\over 2}\right)  [(g^1)^2 + (g^2)^2] \Bigg]
\ ,
\eea
where
\be
K(\tau)= { (\sinh (2\tau) - 2\tau)^{1/3}\over 2^{1/3} \sinh \tau}
\ .
\ee
{}For large $\tau$ we may introduce another radial coordinate $r$ via
\be \label{changeofc}
r^2 = {3\over 2^{5/3}} \varepsilon^{4/3} e^{2\tau/3}\ ,
\ee
and in terms of this radial coordinate
$ ds_6^2 \rightarrow dr^2 + r^2 ds^2_{T^{1,1}}$.

At $\tau=0$ the angular metric degenerates into
\be
d\Omega_3^2= {1\over 2} \varepsilon^{4/3} (2/3)^{1/3}
[ {1\over 2} (g^5)^2 + (g^3)^2 + (g^4)^2 ]
\ ,
\ee
which is the metric of a round $\S^3$ \cite{cd,MT}.
The additional two directions, corresponding to the $\S^2$ fibered
over the $\S^3$, shrink as
\be {1\over 8} \varepsilon^{4/3} (2/3)^{1/3}
\tau^2 [(g^1)^2 + (g^2)^2]
\ .\ee

The simplest ansatz for the 2-form fields is
\begin{eqnarray}
F_3 &=& {M\alpha'\over 2} \left \{g^5\wedge g^3\wedge g^4 + d [ F(\tau)
(g^1\wedge g^3 + g^2\wedge g^4)]\right \} \nonumber \\
&=& {M\alpha'\over 2} \left \{g^5\wedge g^3\wedge g^4 (1- F)
+ g^5\wedge g^1\wedge g^2 F \right. \nonumber \\
&& \qquad \qquad \left. + F' d\tau\wedge
(g^1\wedge g^3 + g^2\wedge g^4) \right \}\ ,
\end{eqnarray}
with $F(0) = 0$ and $F(\infty)=1/2$, and
\be
B_2 = {g_s M \alpha'\over 2} [f(\tau) g^1\wedge g^2
+  k(\tau) g^3\wedge g^4 ]\ ,
\ee
\begin{eqnarray}
H_3 = dB_2 &=& {g_s M \alpha'\over 2} \bigg[
d\tau\wedge (f' g^1\wedge g^2
+  k' g^3\wedge g^4)  
\nonumber \\
&& \left. + {1\over 2} (k-f)
g^5\wedge (g^1\wedge g^3 + g^2\wedge g^4) \right]\ .
\end{eqnarray}

As before, the self-dual 5-form field strength may be
decomposed as $\tilde F_5 = {\cal F}_5 + \star {\cal F}_5$. We
have
\be
{\cal F}_5 = B_2\wedge F_3 = {g_s M^2 (\alpha')^2\over 4} \ell(\tau)
g^1\wedge g^2\wedge g^3\wedge g^4\wedge g^5\ ,
\ee
where
\be
\ell = f(1-F) + k F\ ,
\ee
and
\be
\star {\cal F}_5 = 4 g_s M^2 (\alpha')^2 \varepsilon^{-8/3}
dx^0\wedge dx^1\wedge dx^2\wedge dx^3
\wedge d\tau {\ell(\tau)\over K^2 h^2 \sinh^2 (\tau)}\ .
\ee


\subsection{The First-Order Equations and Their Solution}

In searching for BPS saturated
supergravity backgrounds, the second order equations should be replaced by
a system of first-order ones.
Luckily, this is possible for our ansatz \cite{KS}:
\bea \label{firstorder}
f' &=& (1-F) \tanh^2 (\tau/2)\ , \nonumber \\
k' &=& F \coth^2 (\tau/2)\ , \nonumber \\
F' &=& {1\over 2} (k-f)\ ,
\eea
and
\be \label{firstgrav}
h' = - \alpha {f(1-F) + kF\over K^2 (\tau) \sinh^2 \tau}
\ ,
\ee
where
\be
\alpha =4 (g_s M \alpha')^2
\varepsilon^{-8/3}\ .
\ee
These equations follow from a superpotential for the effective radial
problem \cite{PT}.

Note that the first three of these equations,
(\ref{firstorder}), form a closed system and need to be
solved first.
In fact, these equations imply the self-duality of the
complex 3-form with respect to the metric of the
deformed conifold: $\star_6 G_3 = i G_3$.
The solution is
\bea
F(\tau) &=& {\sinh \tau -\tau\over 2\sinh\tau}\ ,
\nonumber \\
f(\tau) &=& {\tau\coth\tau - 1\over 2\sinh\tau}(\cosh\tau-1) \ ,
\nonumber \\
k(\tau) &=& {\tau\coth\tau - 1\over 2\sinh\tau}(\cosh\tau+1)
\ .
\eea

Now that we have solved for the 3-forms on the deformed conifold,
the warp factor may be determined by integrating
(\ref{firstgrav}).
First we note that
\be
\ell(\tau) = f(1-F) + kF =  {\tau\coth\tau - 1\over 4\sinh^2\tau}
(\sinh 2\tau-2\tau)
\ .\ee
This behaves as $\tau^3$ for small $\tau$.
For large $\tau$ we impose, as usual, the boundary condition that
$h$ vanishes. The resulting integral expression for $h$ is
\be \label{intsol}
h(\tau) = \alpha { 2^{2/3}\over 4} I(\tau) =
(g_s M\alpha')^2 2^{2/3} \varepsilon^{-8/3} I(\tau)\ ,
\ee
where
\be
I(\tau) \equiv
\int_\tau^\infty d x {x\coth x-1\over \sinh^2 x} (\sinh (2x) - 2x)^{1/3}
\ .
\ee
We have not succeeded in evaluating this integral in terms of elementary
or well-known special functions, but it is not hard to see that
\be
I(\tau\to 0) \to a_0 + O(\tau^2) \ ; 
\ee
\be
\ I(\tau\to\infty)\to 3 \cdot 2^{-1/3}
\left (\tau - {1\over 4} \right ) e^{-4\tau/3}
\ ,\ee
where $a_0\approx 0.71805$.
This $I(\tau)$ is nonsingular at the tip of the deformed conifold and, from
\eref{changeofc}, matches the form of the large-$\tau$ solution
\eref{UVs}.  The small $\tau$ behavior follows from the
convergence of the integral \eref{intsol}, while at large
$\tau$ the integrand becomes $\sim xe^{-4x/3}$.

Thus,
for small $\tau$ the ten-dimensional geometry is
approximately $\R{3,1}$ times the
deformed conifold:
\begin{eqnarray} 
ds_{10}^2  &\rightarrow&  { \varepsilon^{4/3}\over
2^{1/3} a_0^{1/2} g_s M\alpha'} dx_n dx_n  +   
a_0^{1/2} 6^{-1/3} (g_s M\alpha')
\bigg \{ {1\over 2} d\tau^2  + {1\over 2} (g^5)^2
\nonumber \\
&&
+ (g^3)^2 + (g^4)^2      + {1\over 4}\tau^2
[(g^1)^2 + (g^2)^2] \bigg \}
\ .
\label{apex}
\end{eqnarray}
This metric will be useful in section 6 where we
investigate various infrared phenomenon of the gauge theory.

Very importantly, for large $g_s M$
the curvatures found in our solution are small
everywhere.
 This is true even far in the IR, since
the radius-squared of the $\S^3$ at $\tau=0$ is of order
$g_s M $ in  string units. This
is the `t Hooft coupling of the gauge theory found far in the IR.
As long as this is large, the curvatures are small and the SUGRA
approximation is reliable.

\subsection{$SO(4)$ invariant expressions for the 3-forms}

In \cite{Grana,Gub} it was shown that the warped background
of the previous section preserves ${\cal N}=1$ SUSY if and only
if
$G_3$ is a $(2,1)$ form on the CY space.
Perhaps the easiest way to see the supersymmetry of
the deformed conifold solution is through a T-duality.
Performing a T-duality along one of the longitudinal directions,
and lifting the result to M-theory maps our background to a Becker-Becker
solution supported by a $G_4$ which is a $(2,2)$ form on
$T^2\times {\rm CY}$. G-flux of this type indeed produces a supersymmetric
background \cite{Becker}.

While writing $G_3$ in terms of the angular 1-forms
$g^i$ is convenient for some purposes, the $(2,1)$ nature of the form
is not manifest. That $G_3$ is indeed $(2,1)$
was demonstrated in \cite{Cvet} with the help
of a holomorphic basis.
Below we write the $G_3$ found in
\cite{KS} in terms of the obvious 1-forms
on the deformed conifold: $dz^i$ and $d\bar z^i$, $i=1,2,3,4$:
\bea
G_3 = \frac{M \alpha'}{2\varepsilon^6 \sinh^4 \tau}
\bigg\{
\frac{\sinh (2\tau)-2\tau}{\sinh \tau}
(\epsilon_{ijkl} \, z_i \bar{z}_j \, dz_k \wedge d\bar{z}_l)
\wedge (\bar{z}_m \, dz_m) \nonumber \\
+ 2 ( 1 - \tau\coth\tau )
(\epsilon_{ijkl} \, z_i \bar{z}_j \, dz_k \wedge dz_l)
\wedge (z_m \, d\bar{z}_m)\bigg \}  .
\eea
We also note that the NS-NS 2-form potential is an
$SO(4)$ invariant $(1,1)$ form:
\be
B_2 = \frac{i g_s M \alpha'}{2 \varepsilon^4}
\frac{\tau \coth \tau - 1}{\sinh^2 \tau}
\epsilon_{ijkl} \, z_i \bar{z}_j \, dz_k \wedge d\bar{z}_l
 \ .
\ee
The derivation of these formulae is given in \cite{remarks}.
Our expressions for the gauge fields are manifestly
$SO(4)$ invariant, and so is the metric.

\section{Infrared Physics}

We have now seen that the deformation of the conifold allows the
solution to be non-singular.
In the following sections we point out some interesting features of the SUGRA
background we have found and show how they realize the expected
phenomena in the dual field theory.  In particular, we will now
demonstrate that there is confinement;
that the theory has glueballs and baryons whose mass scale emerges
through a dimensional transmutation;
that there is a gluino condensate that breaks the $\ZZ_{2M}$ chiral
symmetry down to $\ZZ_2$ and
that there are domain walls
separating inequivalent vacua.
Other stringy approaches to infrared phenomena in ${\cal N}=1$
SYM theory have recently appeared in \cite{MN,Vafa,Tatar}.

\subsection{Dimensional Transmutation and Confinement}

The resolution of the naked singularity via the deformation of the
conifold is a supergravity realization of the dimensional transmutation.
While the singular conifold has no dimensionful parameter, we saw that
turning on the R-R 3-form flux produces
the logarithmic warping of the
KT solution.
The scale necessary to define the
logarithm transmutes into the
the parameter $\varepsilon$ that determines the deformation of the
conifold. From (\ref{changeofc}) we see that $\varepsilon^{2/3}$
has dimensions of length and that
\be
\tau = 3 \ln (r/\varepsilon^{2/3}) + {\rm const}
\ .
\ee
Thus, the scale $r_s$ entering the UV solution
(\ref{UVs}) should be identified with
$\varepsilon^{2/3}$.
On the other hand, the form of the IR metric (\ref{apex})
makes it clear that the dynamically generated
4-d mass scale, which sets the tension of the
confining flux tubes, is
\be
{\varepsilon^{2/3}\over \alpha' \sqrt{ g_s M}
}
\ .
\label{IRscale}
\ee

The reason the theory is confining is
that in the metric for small $\tau$ (\ref{apex}) the function
multiplying $dx_n dx_n$ approaches a constant. This should be
contrasted with the $AdS_5$ metric where this function vanishes at the
horizon, or with the singular metric of \cite{KT} where it blows up.
Consider a Wilson contour positioned at fixed $\tau$, and
calculate the expectation value of the Wilson loop using the
prescription \cite{Juan,Rey}. The minimal area surface bounded by the
contour bends towards smaller $\tau$. If the contour has a very large
area $A$, then most of the minimal surface will drift down into the
region near $\tau=0$. From the fact that the coefficient of $dx_n
dx_n$ is finite at $\tau=0$, we find that a fundamental string with
this surface will have a finite tension, and so the resulting Wilson
loop satisfies the area law.  A simple
estimate shows that the string tension scales as
\be
T_s =
\frac{1}{2^{4/3} a_0^{1/2} \pi}
{\varepsilon^{4/3}\over (\alpha')^2 g_s M } \ .
\label{strten}
\ee
We will return to these confining strings in the next section.

The masses of glueball and Kaluza-Klein (KK) states scale as
\be
m_{glueball} \sim m_{KK} \sim {\varepsilon^{2/3}\over
g_s M \alpha' }
\ .
\ee
Comparing with the string tension, we see that
\be
T_s \sim g_s M (m_{glueball} )^2
\ .
\ee

Due to the deformation, the full SUGRA background has a finite
3-cycle.  We may interpret various branes wrapped over this 3-cycle in
terms of the gauge theory. Note that the 3-cycle has the minimal
volume near $\tau=0$, hence all the wrapped branes will be localized
there. A
wrapped D3-brane plays the role of a baryon vertex which ties
together $M$ fundamental strings.  Note that for $M=0$ the D3-brane
wrapped on the $\S^3$ gave a dibaryon \cite{GK}; the connection between
these two objects becomes clearer when one notes that for $M>0$ the
dibaryon has $M$ uncontracted indices, and therefore joins $M$
external charges. Studying a probe D3-brane in the background of our solution
show that the mass of the baryon scales as
\be
M_b \sim M {\varepsilon^{2/3}\over \alpha'}
\ .
\ee

\subsection{Tensions of the $q$-Strings}

The existence of the blown up 3-cycle with $M$ units of RR
3-form flux through it is responsible for another interesting
infrared phenomenon, the appearance of composite confining strings.
To explain what they are, let us recall that
the basic string corresponds to the Wilson loop in the fundamental 
representation.
The classic criterion for confinement is that this Wilson loop
obey the area law 
\be
-\ln \langle W_1(C) \rangle = T_1 A(C)
\ee
in the limit of large area.  
An interesting generalization is to
consider Wilson loops in antisymmetric tensor representations with $q$
indices where $q$ ranges from $1$ to $M-1$. $q=1$ corresponds to
the fundamental representation as denoted above, and
there is a symmetry under $q\rightarrow M-q$ which corresponds
to replacing quarks by anti-quarks. These Wilson loops can be thought of as
confining strings which connect $q$ probe quarks on one end
to $q$ corresponding probe anti-quarks on the other.
For $q=M$ the probe quarks combine into a colorless state (a baryon);
hence the corresponding Wilson loop does not have an area law.

It is interesting to ask how the tension of this class of confining
strings depends on $q$. If it is a convex function,
\be T_{q+q'} < T_q + T_{q'}
\ ,
\ee
then the $q$-string will not decay into strings with
smaller $q$. This is precisely the situation found by 
Douglas and Shenker (DS) \cite{DS} in softly broken ${\cal N}=2$ gauge theory,
and later by Hanany, Strassler
and Zaffaroni (HSZ) \cite{HSZ} 
in the MQCD approach to confining 
${\cal N}=1$ supersymmetric gauge theory \cite{Wit,MQCD}: 
\be \label{universal}
T_q = \Lambda^2 \sin {\pi q\over M}
\ ,\qquad q=1,2,\ldots, M-1\ 
\ee
where $\Lambda$ is the overall IR scale.

This type of behaviour is also found in the supergravity
duals of ${\cal N}=1$ gauge theories \cite{ChrisIgor}.
Here the confining $q$-string is described by $q$
coincident fundamental strings placed at $\tau=0$ and oriented along the
$\R{3,1}$.\footnote{Qualitatively similar confining flux-tubes were examined in
\cite{CGST} where the authors use the near horizon geometry of
non-extremal D3-branes to model confinement.} 
In the deformed conifold solution analyzed above both $F_5$ and $B_2$
vanish at $\tau=0$, but it is important that there are $M$ units of
$F_3$ flux through the $\S^3$. In fact, this R-R flux blows up the
$q$ fundamental strings into a D3-brane wrapping an $\S^2$ inside the
$\S^3$.
Although the blow-up can be shown directly, for brevity we 
build on a closely related result of Bachas, Douglas and Schweigert \cite{BDS}.
In the S-dual of our type IIB gravity model, 
at $\tau=0$ we find the $\R{3,1}\times \S^3$ geometry with
$M$ units of NS-NS $H_3$ flux through the $\S^3$ and $q$ 
coincident D1-branes along the $\R{3,1}$. T-dualizing along
the D1-brane direction we find
$q$ D0-branes on an $\S^3$ with
$M$ units of NS-NS flux.  This geometry is very
closely related to the setup of
\cite{BDS} whose authors showed that the
$q$ D0-branes blow up into an $\S^2$. 
We will find the same phenomenon, but our probe brane calculation is
somewhat different from \cite{BDS} because the radius of our $\S^3$ is 
different.

After applying S-duality to the KS solution, 
at $\tau=0$ the metric is
\be {\varepsilon^{4/3}\over 
2^{1/3} a_0^{1/2} g_s^2 M\alpha'} dx_n dx_n + b M\alpha' (
d\psi^2 + \sin^2 \psi d\Omega_2^2 )\ ,
\ee
where $b = 2 a_0^{1/2} 6^{-1/3} \approx 0.93266$. 
We are now using the standard round metric on $\S^3$
so that $\psi$ is the azimuthal angle ranging from $0$ to $\pi$.
The NS-NS 2-form field at $\tau=0$ is
\be 
\label{NSfield}
B_2 = M\alpha' \left (\psi -{\sin (2\psi)\over 2} \right)
\sin \theta d\theta\wedge d\phi
\ ,
\ee
while the world volume field is
\be F=-{q\over 2} \sin \theta d\theta\wedge d\phi
\ .
\ee
{}Following \cite{BDS} closely we find that the tension of
a D3-brane which wraps an $\S^2$ located at the azimuthal angle $\psi$ is
\be  
\frac{\epsilon^{4/3}}{12^{1/3} \pi^2 g_s^2 \alpha'^2 b}
\left [ b^2 \sin^4 \psi +
\left (\psi -{\sin (2\psi)\over 2} - {\pi q\over M}\right)^2 \right ]^{1/2}
\ .
\ee
Minimizing with respect to $\psi$ we find
\be \label{trans}
\psi -{\pi q\over M} = {1-b^2\over 2} \sin (2\psi)
\ .
\ee
The tension of the wrapped brane is given in terms of the solution
of this equation by
\be  \label{dbi}
T_q =  \frac{\epsilon^{4/3}}{12^{1/3} \pi^2 g_s^2 \alpha'^2}
\sin \psi \sqrt{ 1 + (b^2 -1) \cos^2 \psi}
\ .
\ee
Note that under $q\rightarrow M-q$, we find $\psi \rightarrow \pi - \psi$,
so that $T_{M-q}= T_q$. This is a crucial property needed for
the connection with the $q$-strings of the gauge theory.

Although (\ref{trans}) is not exactly solvable, we note that
$(1-b^2)/2 \approx  0.06507$ is small numerically. If we ignore
the RHS of this equation, then $\psi\approx \pi q/M$ and
\be \label{probe}
T_q\sim \sin {\pi q\over M}
\ .
\label{confstr}
\ee
The deviations from this formulae are small: even
when $\psi=\pi/4$
and correspondingly $q \approx M/4$, 
the tension in the KS case is approximately $96.7\%$ 
of that in the $b=1$ case.

It is interesting to compare (\ref{confstr}) with
the naive string tension (\ref{strten}) we obtained
in the previous section.  In the large $M$ limit,
we expect interactions among the strings to become
negligible and the $q$-string tension
to become just $q$ times the ordinary
string tension (\ref{strten}).
Indeed, we find that $g_s T_q = q T_s$ in
the large $M$ limit.  The extra
$g_s$ appears because we have been computing
tensions in the dual
background.  When we S-dualize back to
the original background with RR-flux
and $q$ F-strings, all the tensions are
multiplied by $g_s$. 

An analogous calculation for the MN background \cite{MN}
proceeds almost identically. In this background only the $F_3$
flux is present; hence after the S-duality we find only $H_3=dB_2$.
The value of $B_2$ at the minimal radius is again given by
(\ref{NSfield}). There is a subtle difference however from 
the calculation for the KS background
in that now the parameter $b$ entering the
radius of the $\S^3$ is equal to 1. This simplifies the probe calculation
and makes it identical to that of \cite{BDS}. In particular, now we
find
\be \label{mainform}
{T_q\over T_{q'}} = {\sin {\pi q\over M}\over \sin {\pi q'\over M} }
\ ,
\ee
without making any approximations.

Our argument applied to the 
MN background leads very simply to the
DS--HSZ formula for the ratios of $q$-string tensions (\ref{mainform}).
As we have shown earlier, this formula also holds approximately
for the KS background. It is interesting to note that recent lattice
simulations in non-supersymmetric pure glue gauge theory 
\cite{LT} appear to
yield good agreement with (\ref{mainform}).

\subsection{Chiral Symmetry Breaking and Gluino Condensation}

Our $SU(N+M) \times SU(N)$ field theory has an anomaly-free
$\ZZ_{2M}$ R-sym\-metry.
In section 3 we showed that the corresponding symmetry
of the UV (large $\tau$)
limit of the metric is
\be \label{discretesh}
\psi \rightarrow \psi +{2 \pi k\over M}\ , \qquad k=1, 2, \ldots, M
\ .\ee
Recalling that $\psi$ ranges from $0$ to $4\pi$, we see that the full
solution, which depends on $\psi$ through
$\cos \psi$ and $\sin \psi$, has the $\ZZ_2$ symmetry generated by
$\psi \rightarrow \psi + 2\pi$.
As a result, there are $M$ inequivalent vacua:
there are exactly $M$
different discrete orientations of the solution, corresponding to
breaking of the $\ZZ_{2M}$ UV symmetry through the IR effects.
The domain walls constructed out of the wrapped D5-branes separate
these inequivalent vacua.

Let us consider domain walls made of $k$ D5-branes wrapped
over the finite-sized $\S^3$ at $\tau=0$, with remaining directions
parallel to $\R{3,1}$. Such a domain wall is obviously a stable
object in the KS background and crossing it takes us from one ground state
of the theory to another.
Indeed, the wrapped D5-brane produces a discontinuity
in $\int_B F_3$, where $B$ is the cycle dual to the $\S^3$.
If to the left of the domain wall $\int_B F_3=0$, as in the basic solution
derived in the preceding sections, then to the right of the domain
wall
\be
\int_B F_3 = 4\pi^2 \alpha' k
\ ,\ee
as follows from the quantization of the D5-brane charge.
The B-cycle is bounded by a 2-sphere at $\tau=\infty$, hence
$\int_B F_3= \int_{\S^2} \Delta C_2$. Therefore
from (\ref{intforms}) it is clear that
to the right of the wall
\be
\Delta C_2\rightarrow \pi \alpha' k \omega_2
\ee
for large $\tau$.
This change in $C_2$ is produced by the $\ZZ_{2M}$
transformation (\ref{discretesh}) on the original field configuration
(\ref{orig}).

It is expected that flux tubes can
end on these domain walls \cite{dvalishif}. Indeed,
a fundamental string can end on the wrapped D5-brane.
Also, baryons can dissolve in them. By studying a probe D5-brane
in the metric, we find that the domain wall tension is
\be
T_{wall} \sim {1\over g_s} {\varepsilon^2\over (\alpha')^{3}}\ .
\ee

In supersymmetric gluodynamics
the breaking of chiral symmetry is associated with the gluino condensate
$\vev{\lambda\lambda}$. A holographic calculation of the
condensate was carried out by Loewy and Sonnenschein in \cite{SL}
(see also \cite{Bigazzi} for previous work on gluino condensation
in conifold theories.)
They looked for the deviation of the complex 2-form field
$C_2 - {i\over g_s} B_2$ from its asymptotic large $\tau$ form
that enters the KT solution:
\[
\delta \left (C_2 - {i\over g_s} B_2 \right) \sim 
{M\alpha'\over 4} \tau e^{-\tau} [g_1\wedge g_3 + g_2\wedge g_4 -
i( g_1\wedge g_2 - g_3\wedge g_4)]
\]
\be 
\sim  {M\alpha' \varepsilon^2\over r^3}
\ln(r/\varepsilon^{2/3}) e^{i\psi} (d \theta_1 - i\sin \theta_1 d\phi_1)
\wedge  (d \theta_2 - i\sin \theta_2 d\phi_2)
\ .
\ee
In a space-time that approaches $AdS_5$ a perturbation that scales as
$r^{-3}$ corresponds to the expectation value of a dimension 3 operator.
The presence of an extra $\ln(r/\varepsilon^{2/3})$ factor is presumably
due to the fact that the asymptotic KT metric differs from $AdS_5$
by such logarithmic factors. From the angular dependence of the
perturbation we see that the dual operator is $SU(2)\times SU(2)$
invariant and carries R-charge 1.
These are precisely the properties of $\lambda\lambda$.
Thus, the holographic calculation tells us that
\be
\vev{\lambda\lambda} \sim M {\varepsilon^2\over (\alpha')^3}
\ .
\ee
Thus, the parameter $\varepsilon^2$ which enters the deformed
conifold equation has a dual interpretation as the gluino
condensate.\footnote{
It would be nice to understand the relative factor of $g_s M$
between $T_{wall}$ and
$\langle \lambda \lambda \rangle$.
}

\section*{Acknowledgements}
I.~R.~K. is grateful to S.~Gubser, N.~Nekrasov, M.~Strassler,
A.~Tseytlin and E.~Witten for collaboration on parts of the material
reviewed in these notes and for useful input.
This work was supported in part by the NSF grant PHY-9802484.


\end{document}